\documentclass[twocolumn,showpacs,amsmath,amssymb,aps,prl]{revtex4}
\usepackage{graphicx,epsfig}
\def\be{\begin{equation}}
\def\ee{\end{equation}}
\begin{document}

\pagestyle{empty} \preprint{manuscript}

\title{States of Matter }

\author{Deepak Dhar}
\affiliation{Department of Theoretical Physics, Tata
Institute of Fundamental Research, Homi Bhabha Road, Mumbai
400005, India}

\date{\today}

\begin{abstract} This is a written version of a popular science talk for 
school children given on India's National Science Day 2009 at Mumbai. I 
discuss what distinguishes solids, liquids and gases from each other.  I 
discuss briefly granular matter that in some ways behave like solids, 
and in other ways like liquids.

\end{abstract}

\pacs{01.30.Bb, 01.50.Zv, 05.70.Fh}

\maketitle

\section{Introduction}

"It is well-known that matter exists in three forms: solid, liquid and
gas."\\

I picked this quotation from an NCERT textbook for grade XI, but I could 
have picked a different textbook, and found a similar statement there.  
It seems clear and uncontroversial enough, and has been repeated many 
times. What I would like to do in this lecture is to re-examine 
critically this commonly-accepted wisdom. In the process, we will learn 
something about the different states of matter.

  However, telling you more about properties of solids, or liquids, or 
gases is not my main aim. What I really want to do is to impress upon 
you not to accept any such statement without thought. In fact, if you 
bother to think for 5 minutes, all of you can realize that it is not 
fully correct. The student should be able to think for himself/herself, 
and ask if the matter being taught is correct or incorrect. And if it is 
correct, to what extent? So that is my real aim. But we will do it by 
example, by discussing states of matter.

Let us start by asking why is it that there are {\it three} states of 
matter. It is a question like: Why do we live in three-dimensional 
space? Why there are three generations of quarks? and so on.  And with a 
bit of thought, you will realize that the {\it correct} answer in this 
case is that states of matter is a classification scheme, like filing 
cabinets. There are a lot of materials, and we choose the grouping 
scheme that is most convenient. Other possible classification schemes 
could be alphabetical (e.g. in a dictionary), or based on some common 
properties ( zoological classification of different animal species) or 
some mixture of these (e.g. books in a library). Materials can be 
classified based on color, or electrical properties, or whether they are 
organic or inorganic, or conductors or insulators, etc.. All these 
classification schemes are useful, and are used when convenient.

Once we recognize the fact that different states of matter are like 
filing cabinets, the number of cabinets is purely a matter of 
convenience. One can always divide a class into smaller classes, or 
merge smaller classes into a bigger class. So, the {\it number} of 
different states of matter is not a deep question at all: it is whatever 
we want it to be.

Sometimes this sort of discussion ends up being a discussion about 
words. i.e. what is the dictionary definition of solid, liquid and gas?  
We are not discussing words. We are discussing the {\it ideas} behind 
the words. You may say, ``Oh, this is a colloquial word. It doesn't have 
a very precise meaning". But lots of colloquial words have been adapted 
with precise meanings in science. Words like force, work , pressure, in 
ordinary language, can mean a lot of different things. For example, you 
can speak forcefully, or apply political pressure. However, in science, 
the meaning has been restricted to something quite specific, and you can 
quantify it as so many Newtons etc.. So we would expect that even a 
common word like `solid' can be given a specific meaning in science. Can 
we do that? And when we do that, what does it mean?

Let me clarify at the outset that `states'of matter is not the same as 
`phases' of matter. So you can have magnets, for example, and if you 
heat them, they become non-magnetic. There is a {\it phase transition} 
from magnetized phase to non-magnetized phase, but it remains a solid 
throughout. So you have a change of phase but not a change of state.  So 
the `state' of matter is a more general notion than `phases' of matter 
and we are not going to discuss the latter. Also, we will restrict our 
discussion to simple materials. Things like salads, are complicated, and 
not the same everywhere: different parts are different.  We are going to 
discuss only the simpler {\it homogeneous} matter.

The rest of the lecture is organized as follows. First we will discuss 
classification schemes.  Then I will argue why liquid and gases should 
be treated as one state of matter. Then, we will look at differences 
between solids and fluids, and discuss different possible definitions of 
solids. I shall then discuss materials which are different from both 
solids and fluids, and are better treated as a separate state of matter.  
For lack of time, I will discuss only one of these, namely powders, and 
mention some of their unusual properties. And finally summarize our 
discussion.

\section{Requirements of a good classification scheme}

The first requirement of a good classification scheme is that the number 
of classes should be moderate: If you have 500 classes, that is not so 
useful. Each of these classes can be broken up into subclasses, if 
needed. That is what is done with animal classification, and with books 
in a library. But to begin with, you want a moderate number of classes.

Second, there should be a clear and unambiguous definition. If you give 
me an object, I should be able to tell which box it belongs in. There 
should be no confusion whether this is going to be called a solid, or a 
liquid or a gas. It should be a Yes/No answer. No in-betweens.

Thirdly, ease of classification. One should not have to spend a lot of 
effort in trying to decide to which class a given object belongs. You 
will have to do {\it some test} to decide. It is better if the tests do 
not involve expensive, not-easily-available apparatus.

The fourth requirement is naturalness and usefulness: I could have 
listed these separately, but I actually put them together because they 
are, in effect, the same thing. `Natural' means it should not be 
arbitrary or artificial. For example, I can say that something is solid 
if its density if more than $\sqrt{2}$, in some units. This will be an 
unnatural and artificial definition (Why this value?). Artificial and 
unnatural definitions are not useful, as there is a good chance somebody 
else will choose a different number. Then it could be that in India, 
object A is a solid, but not in China. If you make an arbitrary 
definition, it is unlikely to be useful: the object has similar behavior 
whether it is above or below the legal threshold.

The characterization of solids, liquids and gases is known since ancient 
times.  If you are going to the market to buy oil, it helps to know that 
you have to take a bottle along to bring back in. The essential 
difference between solids and liquids is qualitatively how you handle 
them. Can you just pick the matter with bare hands, or do you have to 
use a spoon? The classification in terms of solid, liquid and gas is in 
terms of the {\it mechanical properties} of matter.

\section{Liquids and gases}

\begin{figure}
\begin{center}
\includegraphics[width=4.0cm]{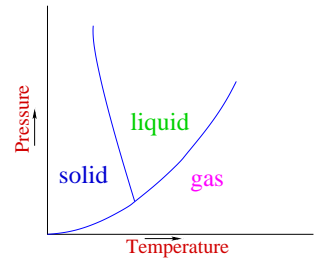}
\caption{Phase diagram of a typical material, showing the solid liquid and gaseous states.}
\end{center}
\end{figure}

Figure 1 gives what is called the phase diagram of a typical material, 
say water. Temperature is plotted along the X-axis and pressure on the 
Y-axis. For a particular temperature and pressure, the matter exists in 
a particular state, marked there tentatively as solid, liquid and gas.  
So, if you fix some pressure and increase the temperature slowly, 
initially you start with solid, which melts to a liquid, and then boils 
to form a gas. However, it is found experimentally that above a 
particular pressure if you heat the liquid, there is no sharp boiling 
point. The material just keeps on getting hotter, without any sudden 
change.  So there is a critical point above which liquid and gas are 
indistinguishable.  If I colour the areas blue, green and red, I will 
have a hard time deciding where to put the boundary between the green 
and red regions. Any choice one makes would be arbitrary.  In other 
words, liquids and gases are the same state. And together, they are 
called the {\it fluid state}. This is actually well-known, but the 
school textbooks continue to preach that there are three states of 
matter.

Thus, I have reduced my problem from three states of matter to two. 
Sometimes people say that plasma is the fourth state of matter. What is 
plasma? If you take a gas and heat it, more and more of the atoms become 
ionized as the temperature increases. When the material is very heavily 
ionized, it is called `plasma'.  Plasmas respond strongly to electric 
fields. However, again, between gas and plasma, there is no sharp point 
of transition. Ionization increases smoothly as you increase 
temperature. So any boundary between gas and plasma, will be an 
arbitrary boundary, and not a natural distinction. And hence plasma and 
gaseous state are not distinct: they are the same state.

\section{Solids and fluids}

Let us now see how we can distinguish solids from fluids.

{\underline {Def. 1}}:``Solids have a fixed shape and fixed volume but 
liquids have a fixed volume but no fixed shape, while gases have neither 
fixed shape nor fixed volume."

\begin{figure}
\begin{center}
\includegraphics[width=8.0cm]{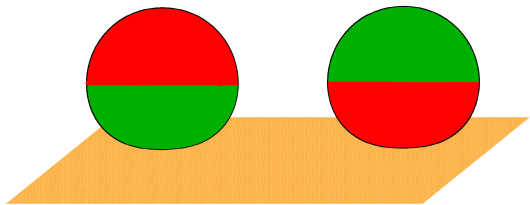}
\caption{solid balls will deform under gravity}
\end{center}
\end{figure}

This is a definition that I remember from my schooldays. Perhaps it is 
in your textbook also. But let us look at it more closely. In Fig. 2, I 
have shown two pictures of a ball ( made of say rubber, half colored 
red, half green): Clearly, the ball gets a little squashed when you put 
it on the table and if you flip it, it is again squashed, but the shape 
is changed!.  So, is it a solid or not a solid? In fact, all materials 
have finite compressibility, and will deform somewhat under force.

Let us try an alternate  definition:

{\underline {Def. 2}}:``In solids, the atoms vibrate about their mean 
positions, but in fluids, they move over all available space.''

This poses a bit of a problem because it refers to molecules I cannot 
see with the naked eye. But if you take an atom in a solid, mark it in 
some way and observe its movement, then one would see, over time that it 
actually jiggles over larger and larger distances. So all particles 
diffuse in time and this diffusion constant is finite.  The mean square 
displacement $\langle R^2\rangle$ of a tagged particle is expected to 
linearly increase with time : $\langle R^2 \rangle \sim Dt$.

The diffusion constant $D$ is finite. In solids, an atom moves about 
0.0001 mm in one minute. In liquids about 1 mm about 100 mm in gases. So 
it is not correct to say that, in solids, atoms do not move over all 
available space. They would, if you wait long enough.

 Let us try  another definition: 

{\underline {Def. 3}}:``Solids have a long range ordered periodic 
arrangement of atoms. Fluids have only a short range order.''

\begin{figure}
\begin{center}
\includegraphics[width=6.0cm]{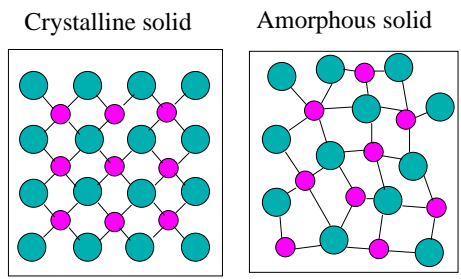}
\caption{ Schematic representation of the atomic arrangements in 
crystalline and amorphous materials}
\end{center}
\end{figure}

In solids you expect a regular, periodic arrangement and in liquids, you 
have an irregular arrangement. So first, there is thermal motion in 
atoms; they are not in fixed position, they are jiggling around. Any time you 
take a snapshot, you will not find this nice periodic arrangement shown 
above. Each atom will be slightly displaced. So how does one distinguish 
this from the rest? There is a technical way of distinguishing it which 
says that take an x-ray diffraction picture of this. If you get sharp 
peaks, then there is an overall periodic structure. If there are no 
sharp peaks, then it is not a long-range periodic ordered structure. 

However, if you say that long range periodic ordered structures are 
solids and other are not, plastic or some amorphous material like window 
glass, will not qualify as solids. So this definition is also not a good 
definition. There is another definition which says:

{\underline {Def. 4}}:``Solids have a finite shear modulus, liquids do not."

Maybe you have not seen that definition above but that is the one most 
physicist like the most. What is the shear modulus? Suspend a weight 
from a wire from the ceiling.  if you apply a twist to the weight, and 
let go, it starts to oscillate, and you get a torsion pendulum.  If 
there was no restoring force which tries to undo the twist, it would not 
oscillate. Now suppose you have two cylinderical pipes one inside the 
other, you fill some liquid between these and then you apply a twist to 
the inner pipe. In this case, there is no restoring force. So there is 
no restoring force to shear in liquids, but it is in solids. This is the 
usually accepted distinction between solids and liquids.

However, it also has a problem because, what happens is that if you take 
a solid and you apply this twist, and you hold it for a long time, then 
the force felt by the rod slowly decreases with time.  This is called 
creep in solids. The material re-adjusts under this strain and the 
molecules move to relieve this strain. Thus how much shear modulus or 
how much restoring force there is, depends on how much time you wait 
before you measure it. And so if you take a really long time then may be 
it does not feel any force.

The definition of shear modulus involves very slowly changing forces, 
and it would appear that if you really wait very long, the shear moduls 
is always zero.

I was looking up other possible definitions for distinctions between 
solids and liquids and there is one which is not so often used in school 
textbooks but it was used as a distinguishing characteristic for 
materials in our ancient books.  
(\includegraphics[width=4.0cm]{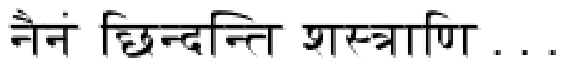} ).

{\underline {Def. 5}}:``Solids can be cut with a knife, fluids not''.\\

This also looks like a reasonable characterization.  A tougher solid is 
harder to cut, but eventually you can cut it. For a liquid like water, 
there is no use trying to cut it with a knife. Even this definition 
turns out to be not very useful because there are things like cold 
welding of solids. You can take two solids, you can cut them, put them 
on top of each other, vibrate them a little bit and they become the same 
again. This is called cold welding.  In the liquid, you cut it with a 
knife and the two separated parts behind the knife's moving edge re-join 
again. This self-healing after the cut can occur in solids as well as 
liquids.

This is becoming a bit confusing. To recall, we started by saying that 
one of the requirements for a good classification scheme was a clear, 
unambiguous definition. It seems reasonable to expect, but now, we find 
this difficult to satisfy.

In fact, {\it{`solid-like' and `fluid-like' behavior is a matter of 
length and time scales.}} A small drop of water or mercury resists 
change of shape, and is quite "rigid". A "ductile" metal flows at 
long-enough time scales. Falling from a plane on a lake, you are likely 
to break your bones, as badly as falling on hard ground. A single brick 
is clearly a solid, a truck-load of bricks can be poured out, and 
acquires fluid-like properties.

\section{Powders: a different state of matter}

Powders are granular materials like sand, wheat, flour. I would now like 
to argue that they are a state of matter different from both solids and 
fluids. Firstly, one can broadly distinguish between two types: wet and 
dry powders.  In the following, I restrict myself to the former.
  
 Firstly, powders can be poured from one vessel to another, and take the 
shape of the vessel. In this sense, they are like fluids. However, if 
pour a powder on a flat table, from a point above, they form a conical 
pile, in which the slanting surface makes a finite angle with the 
horizontal.  This angle is characteristic of the material, and is called 
the angle of repose. If powders were fluid, this angle would have been 
zero.

\begin{figure}
\begin{center}
\includegraphics[width=6.0cm]{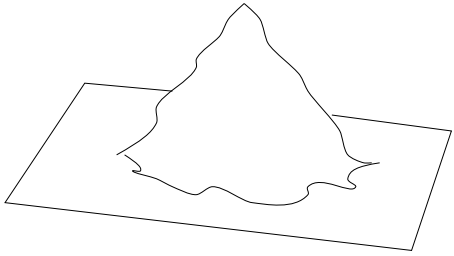}
\caption{A small sandpile on a flat table} 
\end{center}
\end{figure}

In general, the flow of powders is very different from fluids. While the 
latter is quite well understood, and is described by laws of 
hydrodynamics, even the equivalent of hydrostatics for powders is only 
partly understood. For example,  consider a long cylindrical 
vessel, which is filled with sand or water up to a height $h$. In the 
case of fluids, the pressure at the base of the vessel increases 
linearly with the height of the column. In case of sand, it initially 
increases, but tends to a finite saturation value even as the height is 
increased [Fig. 5].

\begin{figure}
\begin{center}
\includegraphics[width=5.0cm]{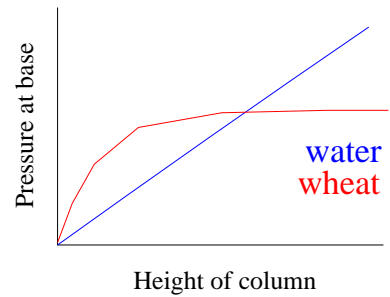}
\caption{Pressure at base of a cylinder filled with water and sand, as  a function of height of the column}
\end{center}
\end{figure}

Another interesting behavior of powders is called the ``brazil-nut 
effect". This refers to the fact that larger heavy particles rise to top 
in shaken granular media. The name refers to the well-known phenomena of 
larger nuts being found near the top in a shaken cereal box. I 
understand that the same phenomena can be seen with large cannonballs in 
a shaken box of sand. In liquids, the heavier cannonballs would be 
expected to sink, not rise. Clearly, the well-known law of Archimedes is 
not valid for powders.

 This coming together of larger particles in shaken mixtures has 
important consequences. In many applications, it is important for 
powders to be well-mixed, e.g. in making medicine tablets. But, if size 
separation occurs, keeping a well-mixed granular mixture in a rotating 
drum, would make it unmixed.

\section{Other  states of matter}

   There are many other types of matter whose behaviour differs 
substantially from solids and fluids, so that it seems reasonable to 
classify them as separate states of matter. A detailed discussion of 
each of these would require too much time. I only mention them briefly 
here. You can learn more about them from books and from the internet.

   For example, glasses, like window glass, looks like a solid, but 
their atomic structure is like that of liquids, and they seem to flow, 
though very slowly. They can be thought of as very viscous liquids, but 
perhaps better thought of as a distinct state of matter. Liquid 
crystals, the stuff used in your mobile phone displays, flow like 
fluids, but show partly crystalline atomic order, and can be thought of 
as a state of matter between solids and fluids. These are 'solid -like' 
at atomic level, but 'liquid-like' in bulk behavior, while in glasses, 
the converse happens. Helium at low temperatures when it becomes a {\it 
superfluid}, or Bose -Einstein condensates are rather exotic forms of 
matter, not encountered in everyday life, but they have very unusual 
flow properties and certainly qualify as distinct states of matter. 
Recently, there has been some indication of experimental evidence of a 
state called `supersolids', that have periodic atomic arrangement in 
space, but flow like superfluids. Then there are colloids, gels, 
emulsions, foams $\ldots$. These could also be considered as different 
states of matter, but one can argue that they are not really 
homogeneous. Away from earth, in neutron stars, one has matter in the 
form of neutrons, and that is ceratinly a new state of matter. 
Astrophysicists these days even speak of `dark matter',  which, if it 
is found to exist,   is going to be  very 
different from other known forms of matter.

\section{Summary}

 To summarize, the main thing I have tried to emphasize in this talk is 
the importance of thinking for yourself in whatever you study. Another 
is that we should correct our textbooks. Sometimes, the changes required 
are not so large, and some textbooks do say things correctly. For 
example, the textbook ``Advanced Chemistry'' by P Mathew, (Cambridge 
Univ. Press) says: "Almost all substances fall neatly into one of the 
three categories: solid, liquid and gas $\ldots$".

  I argued that it is difficult to give a clear precise definition of 
different states of matter: It is a question of length and time scales.  
The surface of water is hard, solid-like for large velocity impact, and 
ductile metals flow (can be pulled into wires).  Powders are examples of 
states of matter different from both solids and fluids.
 
 And, finally, the students should realize that there is much we do not 
understand, even about everyday life objects. Understanding things 
better can be very exciting.  It is good to remember that today, on our 
National Science Day.

\end{document}